\documentstyle[twoside,preprint,pra,aps]{revtex}
\tighten
\begin{document}
\preprint{\today}
\draft
%
%%%%%%%%%%%%%%%%%%%%%%%%%%%%%%%%% TITLE PAGE
%
\title{Impulsive quantum measurements: \\
restricted path integral versus von Neumann collapse}
\author{Tommaso Calarco}
\address{Dipartimento di Fisica, 
Universit\`a di Ferrara,\\
Via Paradiso 12, Ferrara, Italy 44100}
\date{\today}
\maketitle
%
%%%%%%%%%%%%%%%%%%%%%%%%%%%%%%%%% ABSTRACT
%
\begin{abstract}
The relation between the restricted path integral approach to quantum 
measurement theory and the commonly accepted von Neumann wavefunction 
collapse postulate is presented. 
It is argued that in the limit of impulsive measurements the two 
approaches lead to the same predictions. 
The example of repeated impulsive quantum measurements of position 
performed on a harmonic oscillator is discussed in detail and the quantum 
nondemolition strategies are recovered in both the approaches.
\end{abstract}
%
%%%%%%%%%%%%%%%%%%%%%%%%%%%%%%%%% PACS NUMBERS
%

\narrowtext

\begin{section}{Introduction}

Despite the impressive successes of quantum mechanics in explaining most 
experimental results about microscopic phenomena, a unique consistent quantum 
theory of measurement is still missing. In this field, an important achievement 
was made by von Neumann which postulated two ways for the evolution of the 
state vector: continuously, according to the linear Schr\"{o}dinger 
equation, when time passes without a measurement on the system being performed;
and discontinuously, according to probability laws, if a measurement is carried
out (the so-called wavefunction collapse \cite{vN}). 
von Neumann first defined this feature ``a peculiar dual nature of the 
quantum mechanical procedure, which could not be satisfactorily 
explained"\footnote{{\sl ``eine eigenartige Duplizit\"{a}t des Vorgehens, die 
nicht gen\"{u}gend erkl\"{a}rt werden k\"{o}nnte"} \cite{vN} (p.~222 in the
original version; p.~417 in the English translation).}.\\
Mensky proposed \cite{ME} a phenomenological theory, based on Feynman path
integral formalism \cite{FEYN}, that expresses the {\em a
posteriori} dynamical evolution of a system, undergoing a
continuous measurement, in terms of the instrumental uncertainty and the
output of the meter, supposed to be known {\em before} doing the calculations.
The effect of the measurement is introduced in the space of the paths by means
of an influence functional, which restricts the integration to those paths that
lie around the measurement result. The restricted path integral approach has
been applied to describe continuous \cite{MEOP1} and impulsive \cite{MEOP2}
measurements in both linear and non-linear systems, to explain quantum
Zeno effect \cite{OPTA} and to verify the possibility of testing quantum
mechanics through temporal Bell-like inequalities in bistable potentials
\cite{CALARCO}.\\
The most common objection raised against this approach is that it seems to
treat the Feynman paths, in some sense, as real trajectories followed by the
system. Mathematically, this means that the time evolution during the
measurement is non-unitary, {\em i.e.} the wavefunction looses its
normalization. In fact, it has been shown \cite{MEOP1} that the obtained
propagator is the same as the one associated to an effective Hamiltonian having 
a purely imaginary measurement term, which of course destroys the unitarity.

The aim of the present paper is to show how this effect is 
equivalent, in the impulsive limit, to von 
Neumann collapse. Section II is devoted to derive, from the restricted
path integral propagator for a measurement of infinitesimal
duration performed on a generalized coordinate, an analytic formula with which 
von Neumann formalism can be
recovered. In Section III the calculation for a series of impulsive 
measurements in a generic potential with discrete energy levels, only requiring 
the knowledge of the energy eigenstates,
is developed and is applied --~in Section IV~-- to 
the case of repeated impulsive measurements of position in a harmonic 
oscillator \cite{MEOP2}. 
Finally the predictions of Mensky and von Neumann theories are compared and 
conclusions are drawn in Section V.

\end{section}

\begin{section}{Quantum model for impulsive measurements of position}

The path integral formulation of quantum mechanics \cite{FEYN} provides a 
natural framework for handling continuous quantum measurements of position by
restricting the integration, in the space of the trajectories $x(t)$ in the 
coordinate $x$  --~assumed to be continuously monitored between times 0 and 
$\tau$~--, to those paths which turn out to be compatible with the experimental 
outcome $a(t)$ within the instrumental error $\Delta a$ \cite{ME}. 
This can be done by means of a weight functional $w_{[a]}[x]$ depressing the 
contribution of paths whose distance from the $a(t)$ actually obtained (which 
is in general a continuous but not necessarily differentiable function of time)
exceeds $\Delta a$. 
The propagator for the system is then written as a weighted integral:
\begin{equation}
\label{PROPK}
K_{[a]}(x^{\prime\prime},\tau;x^{\prime},0)=
\int_{x(0)\equiv x^{\prime}}^{x(\tau)\equiv x^{\prime\prime}}
 d[x]\exp \left\{\frac{i}{\hbar}\int_0^{\tau}{\cal L}(x(t),
\dot{x}(t),t)dt \right\}w_{[a]}[x].
\end{equation}
The probability distribution for the measurement output is a 
functional of $a(t)$; if $\psi(x,0)$ is the wavefunction representing the 
initial state of the system, it has the form
\begin{equation}
\label{PGEN}
P_{[a]}=\frac{\vert\langle\psi_{[a]}(\tau)|\psi_{[a]}
(\tau)\rangle\vert^2}{\int\vert\langle\psi_{[a]}(\tau)|\psi_{[a]}(\tau)\rangle
\vert^2 d[a]},
\label{P}
\end{equation}
where 
\begin{equation}
\psi_{[a]}(x^{\prime\prime},\tau)={\int} 
K_{[a]}(x^{\prime\prime},\tau;
x^{\prime},0)\psi(x^{\prime},0)dx^{\prime}.
\label{PSITAU}
\end{equation}
Its dispersion estimates the actual experimental accuracy with which it is 
possible to extract the information on the position $x$, also called effective 
uncertainty: 
\begin{equation}
\Delta a_{\rm eff}^2=
2 \int \frac{1}{\tau} \int_0^{\tau} [a(t)-\tilde{a}(t)]^2 dt~ 
P_{[a]}d[a]. 
\label{DAEFFTH}
\end{equation}
\noindent
where $\tilde{a}(t)$ is the path which maximizes $P_{[a]}$. Of course, in
general $\Delta a_{\rm eff}\geq\Delta a$.

The most natural way to represent in this framework
an impulsive measurement at the initial instant is to consider it 
as the limit for infinitesimal time intervals of a continuous one 
with constant result $a(t)\equiv a$. Of course, in this case 
the probability distribution for the measurement results is a function of $a$.

Alternatively, one can take the limit directly in the path integral 
expression~(\ref{PROPK}). A simple form for the weight functional is the 
Gaussian one \cite{MEOP1}:
\begin{equation}
\label{PESOW}
w_{[a]}[x]=\exp \left\{-\kappa\int_0^{\tau} [x(t)-a(t)]^2 dt \right\},
\end{equation}
where the measurement coupling $\kappa$ in general should be taken 
constant, in order to ensure that the dynamics 
can be described by a semigroup \cite{ME}. In the limit 
$\tau\to 0$, it turns out to be useful the position
\begin{equation}
\label{KAPPA}
\kappa=\frac{1}{2\Delta a^2\tau},
\end{equation}
in which $\Delta a$ assumes the proper meaning of a width in the space of
paths. In this way, regardless of the form of the Lagrangian involved, one 
obtains an analytical expression \cite{CALARCO}:
\begin{eqnarray}
\label{PROPDEL}
K_{a}(x^{\prime\prime},x^{\prime}) & = & \lim_{\tau\to 0}
K_{[a(t)\equiv a]}(x^{\prime\prime},\tau;x^{\prime},0)=\nonumber\nopagebreak\\
& = &  \lim_{\tau\to 0}\int_{x(0)\equiv x^{\prime}}
^{x(\tau)\equiv x^{\prime\prime}}
d[x]\exp \left\{\int_0^{\tau}\left(\frac{i}{\hbar}{\cal L}-
\frac{[x(t)-a]^2}{2\Delta a^2 \tau}\right)dt\right\}=\nonumber\nopagebreak\\
& = & \lim_{\tau\to 0}
\int_{x(0)\equiv x^{\prime}}
^{x(\tau)\equiv x^{\prime\prime}}
d[x]\exp \left\{-\frac{[x(0)-a]^2\not\tau}
{2\Delta a^2 \not\tau}\right\}=\nopagebreak\\
& = & e^{-\frac{(x^{\prime}-a)^2}{2\Delta a^2}}
K(x^{\prime\prime},0;x^{\prime},0)\equiv\nonumber\nopagebreak\\
& \equiv & e^{-\frac{(x^{\prime}-a)^2}{2\Delta a^2}}\delta(x^{\prime
\prime}-x^{\prime}).\nonumber
\end{eqnarray}
The third line follows from neglecting, in the limit, the Lagrangian term with
respect to the measurement term, going as $\tau^{-1}$ (this
appears reasonable also from the physical point of view, because an impulsive
measurement is assumed to induce a significant change in the state of the
system during a negligible amount of time), and by applying the theorem of the 
mean value. $K$ is the propagator in the absence of measurement.\\
Let $\psi(x,t)$ be the wavefunction which describes a system undergoing 
an impulsive measurement at the instant $t$, with result $a$.
From Eqn.~(\ref{PROPDEL}) follows
\begin{equation}
\psi_a(x,t^+)=w_a(x)\psi(x,t^-), 
\end{equation}
where
\begin{equation}
\label{wepsi}
w_a(x)\stackrel{\rm def}{\equiv}e^{-\frac{(x-a)^2}{2\Delta a^2}}.
\end{equation}
Thus $\parallel\psi_a(t^+)\parallel$ is the projection ({\em
i.e.} the scalar product modulus)
of $\psi(x,t^-)$ on the weight function $w_{a}(x)$. So the quantity
\begin{equation}
\label{pepsi}
P(a) = \frac{\parallel\psi_a(t^+)\parallel^2}{
\int\!\!\int e^{-\frac{(x-a)^2}{\Delta a^2}}|\psi(x,t^-)|^2dx\, da}=
\frac{1}{\sqrt{\pi}{\Delta a}}\parallel\psi_a(t^+)\parallel^2
\end{equation}
represents --~in analogy with Eqn.~(\ref{P})~-- the 
probability that the system is found in the state described by
$w_{a}(x)$.\\
It should be noted that in the limit $\tau\to 0$, if $\Delta a$ remains 
finite, $\kappa$ diverges as $\tau^{-1}$: this means that von Neumann collapse 
is recovered by considering not only infinitesimal measurement durations, but
also infinite coupling between the instrument and the system. Furthermore, in 
the limit of an extremely precise measurement ($\Delta a\to 0$), 
$\kappa$ diverges more rapidly than $\tau^{-1}$, and one easily obtains
\begin{equation}
\lim_{\Delta a\to 0}P(a)=|\psi(a,t^-)|^2
\end{equation}
in agreement with the conventional interpretation of the
wavefunction.\\
It is worth noting also that $\parallel\psi_a(t^+)\parallel^2$ can  be less than
1: this follows from the non-unitarity of the temporal evolution induced in
Eqn.~(\ref{PROPK}) by the measurement term. Thus, in the impulsive limit, that 
apparently arbitrary modification of the dynamics of the system under monitoring
is equivalent to the commonly accepted von Neumann postulate \cite{vN} of the 
discontinuity introduced by the measurement in the causal linear evolution of 
the state. For instance, the case of a perfect measurement
which yields with certainty the information whether a particle is found within
an interval of width $\Delta a$ around the position $a$, is
recovered by choosing 
for the measurement operator the form, discontinuous and therefore 
less realistic than the (\ref{wepsi}),
\begin{equation}
\label{misvn}
\hat{w}^{v.N.}_a\propto 
\theta(\hat{x}-[a-\Delta a])\theta([a+\Delta a]-\hat{x}).
\end{equation}
The restricted path integral approach to continuous measurements with a generic
weight functional $w_{[a]}[x]$ appears therefore simply as a generalization of 
the idea of perfect instantaneous filtering in von Neumann theory of
measurement being a smoothed version of it with finite accuracy and
duration. The comparison between the results of the two approaches will be
now performed in the case of stroboscopic measurements of position on a 
harmonic oscillator \cite{MEOP2}.
\end{section}

\begin{section}{Stroboscopic measurements of position}

A stroboscopic sequence of measurements is obtained when an observable is
monitored in an impulsive way at some definite instants,
equally spaced by a quiescent time $\Delta T$ in which no measurement is 
performed. Such a topic has been studied in detail \cite{MEOP2} for  
characterizing Quantum Non Demolition \cite{CAVES} strategies for the 
measurement of the position of a quantum system. 
Particular advantages, in this field,
can arise from applying the method developed in the preceding section and 
exploiting an energy eigenstates expansion, because of the uniformity to 
handle each form of the weight function $w_a(x)$,
for instance the na\"{\i}ve one (\ref{misvn}) --~which expresses the usual 
representation of the measurement~--, allowing analytical calculations for 
every potential \cite{CALARCO}. For a generic system having discrete energetic 
levels ({\em i.e.} $H|l\rangle=E_l|l\rangle$),
an initial state can be developed in energy eigenstates:
\begin{equation}
\label{SVILPSI}
|\psi(t_0)\rangle=\sum_{l=1}^{\infty}c_l|l\rangle.
\end{equation}
If the coordinate of the system 
is measured with results $a_n$, $n=0,1,\ldots,N$ at each of 
the instants $t_n\equiv n\Delta T$, we get
\begin{equation}
\label{PSIFIN}
|\psi_{\{a_n\}_{n=0,\ldots,N}}(t_N^+)\rangle=\hat{w}_{a_N}
\left(\prod_{j=1}^N
e^{-\frac{i}{\hbar}\hat{H}\Delta T}\hat{w}_{a_{N-j}}\right)
|\psi(t_0^-)\rangle,
\end{equation}
where $\hat{w}_a$ is the multiplication operator corresponding to the weight 
function (\ref{wepsi}).
The normalization constants relative to each measurement are factorized and 
therefore can be neglected in the 
calculations: because they simplify in the definition (\ref{P}) of $P(a)$,  
$\Delta a_{\rm eff}$ will not depend on them.

By considering Eqn.~(\ref{SVILPSI}) and by inserting a completeness 
${\bf 1}\equiv\sum_{m=1}^\infty |m\rangle\langle m|$,
Eqn.~(\ref{PSIFIN}) is rewritten
\begin{eqnarray}
\label{PSIN}
|\psi_{\{a_n\}_{n=0,\ldots,N}}(t_N^+)\rangle & = & 
\sum_{l,m=1}^\infty c_l
|m\rangle\langle m|\hat{w}_{a_N}e^{-\frac{i}{\hbar}\hat{H}\Delta T}
\hat{w}_{a_{N-1}}\cdots\hat{w}_{a_1}e^{-\frac{i}{\hbar}\hat{H}\Delta T}
\hat{w}_{a_0}|l\rangle=\nonumber\\
& = & \sum_{m=1}^\infty \left(\sum_{l=1}^\infty B_{ml}^Nc_l\right)
|m\rangle,
\end{eqnarray}
where
\begin{equation}
B_{ml}^N(\Delta T,\Delta a,\{a_n\})\stackrel{\rm def}{=}\langle m|
\hat{w}_{a_N}\left(\prod_{j=1}^N
e^{-\frac{i}{\hbar}\hat{H}\Delta T}\hat{w}_{a_{N-j}}\right)|l\rangle.
\end{equation}
By inserting $N$ times the identity operator
${\bf 1}\equiv\sum_{n_i=1}^\infty |{n_i}\rangle\langle {n_i}|$
($i=1,2,\ldots,N$), one gets
\begin{equation}
\label{BML}
B_{ml}^N=\sum_{n_1,n_2,\ldots,n_N=1}^\infty 
W_{mn_1}^{a_N}\left(\prod_{j=1}^{N-1}W^{a_{N-j}
}_{n_jn_{j+1}}\right)
W_{n_Nl}^{a_0}\exp\left\{-\frac{i\Delta T}{\hbar}
\sum_{i=1}^{N}E_{n_i}\right\},
\end{equation}
with
\begin{equation}
\label{WIJ}
W_{ij}^{a_n}(\Delta a)\stackrel{\rm def}{\equiv}\int_{-\infty}^{+\infty}
u_i^\ast(x) w_{a_n}(x)u_j(x)dx,
\end{equation}
in which the $u_i(x)\equiv\langle x|i\rangle$ are the 
energy eigenfunctions in $\{{\cal X}\}$-representation. 

Let $P_{a_0,a_1,\ldots,a_{N-1}}(a)$ be the probability that the $N^{\rm th}$ 
measurement will give result $a$, when the results of all the previous 
measurements are known. Its dispersion can be now evaluated, in analogy
with Eqn.~(\ref{DAEFFTH}), from Eqn.~(\ref{PSIN}) through 
Eqs.~(\ref{BML})-(\ref{WIJ}). It is no more necessary to take the mean value 
on the measurement time as in (\ref{DAEFFTH}), since $\tau\to 0$. The final 
expression is \cite{CALARCO}:
\begin{equation}
\label{DPHIEFF}
\Delta a_{\rm eff}^2(\{a_n\}_{n\leq N-1},N)=
2\int_{-\infty}^{+\infty}(a-\tilde{a}_N)^2\left(\sum_{m=1}^\infty
\left|\sum_{l=1}^\infty B_{ml}^N(a_0,\ldots,a_{N-1},a)
c_l\right|^2\right)^2da,
\end{equation}
where $\tilde{a}_N$ is the most probable result of the $N^{\rm th}$
measurement, and the explicit dependences of $\Delta a_{\rm eff}$ 
and of $B_{ml}^N$ from $\Delta T$ and from $\Delta a$ have been omitted.

Summarizing, once eigenvalues and eigenstates 
of a generic quantum system with discrete energy spectrum are known, 
from the decomposition in eigenstates of the initial wavefunction it
is possible in principle to infer directly the value of the effective 
uncertainty after an arbitrary sequence of stroboscopic measurements.
The actual calculation is difficult due to the presence of multiple 
sums on the numerable ensemble of the energy eigenstates. 
A practical evaluation requires to approximate the sums and the integrals in 
Eqn.~(\ref{DPHIEFF}),
by truncating them to finite values, say respectively $N_{\rm MAX}$ and $\pm
A_{\rm MAX}$. The accuracy of such an approximation has to be checked for
comparison with results already known by other methods, both analytical (if any)
or numerical.
\end{section}

\begin{section}{Optimal measurements of position for a harmonic oscillator}

The harmonic potential shows its great importance in many fundamental problems 
in physics. In particular, it has been studied in the limit of infinite 
coupling for impulsive measurements, to search for Quantum Non Demolition (QND) 
observables \cite{CAVES}. These last are relevant for the detection of small 
displacements in
the quantum limit of sensitivity for mechanical resonators used as 
gravitational wave antennas. In the present framework this problem can be 
formulated by noting that in general the effective uncertainty 
$\Delta a_{\rm eff}$ is greater than the instrumental error $\Delta a$, 
expressing the spreading of the paths due to the effect of the back-action 
of the meter on the measured system. However for certain observables --~namely 
the QND ones~--, the ratio of the two uncertainties can be reduced to unity by 
applying optimal measurement strategies, without violating  
the Heisenberg principle \cite{MEOP2}. In this section the method described 
before will be applied to the characterization of such QND strategies for a 
harmonic oscillator. The results obtained through the two measurement theories 
under consideration, expressed respectively by the weight functionals 
(\ref{PESOW}) and (\ref{wepsi}), will finally be compared.

For a harmonic oscillator, described by a Lagrangian of the form
${\cal L}=\frac 12 m\dot{x}^2-\frac 12 m\omega^2x^2$, it is possible to
characterize also {\em a priori} the QND strategies for stroboscopic
measurements of position, by means of the commutation relation of the observable
$\hat{x}_{\rm H}$ --~in the Heisenberg picture~-- at different times \cite{CAVES}:
\begin{equation}
\label{COMM}
[\hat{x}_{\rm H}(t),\hat{x}_{\rm H}(t+\Delta T)]=\frac{i\hbar}{m\omega}
\sin(\omega\Delta T).
\end{equation}
From Eqn.~(\ref{COMM}) follows that a series of impulsive measurements of 
position with infinite precision, performed every $T/2$ --~where $T=2\pi/\omega$
is the oscillation period~--, will
give perfectly predictable outcomes. Therefore one expects that, for impulsive 
measurements spaced in such a way, the inequality $\Delta a_{\rm
eff}\geq\Delta a$ could be saturated. 
This can be seen also by looking to the periodical wavepacket reformation via
causal dynamical evolution. In the absence of measurements, the wavefunction
$\psi(x,t_0)\equiv\langle x|\psi(t_0)\rangle$ of the state (\ref{SVILPSI})
will be, after half an oscillation period --~in the same notations as
above, specialized to the harmonic potential~--,
\begin{eqnarray}
\psi\left(x,t=t_0+\frac{T}{2}\right)& = &
e^{-\frac{i}{\hbar}\hat{H}(t-t_0)}\sum_{l=1}^{\infty}c_l\,u_l(x)=\nonumber\\
& = & \sum_{l=1}^{\infty}e^{-\frac{i}{\hbar}\left(l+\frac
12\right)\hbar\omega\frac{T}{2}}c_l\,u_l(x)=\\
& = & e^{-i\frac{\pi}{2}}\sum_{l=1}^{\infty}e^{-il\pi}c_l\,(-1)^lu_l(-x)=
\nonumber\\
& = & -i\psi(-x,t_0),\nonumber
\end{eqnarray}
the third line following from the definition of $T$ and from the symmetry
properties of the energy eigenfunctions.\\
Thus, every half-period, the wavepacket is reformed 
symmetrically with respect to the equilibrium position, except for an 
irrelevant phase factor. This means that, if $\psi(x,t_0)$ represented the
state after an impulsive, infinite-precision ($\Delta a\to 0$) measurement with 
outcome $a_0$, another measurement at $t=t_0+T/2$ will yield with certainty the 
result $-a_0$; at $t=t_0+T$ this will be again $a_0$. Therefore, in the
case $a_0=0$ \cite{MEOP2}, the optimal QND strategy 
--~in which $\Delta a_{\rm eff}\to\Delta a$, also for $\Delta a>0$~-- is 
obtained by choosing the quiescent time $\Delta T=T/2$. On the other hand, if 
$a_0\not=0$, the optimality 
is reached either with $\Delta T=T/2$ and alternated results $a_n=(-1)^na_0$
($n=0,1,2,\ldots$), or with constant results $a_n\equiv a_0$ but $\Delta
T=T$.\\
Indeed, this behavior has been already obtained \cite{MEOP2} by simulations 
based on the restricted path integral; here it will be recovered via von Neumann
collapse, by means of the calculation described 
in the previous section. A preliminar comparison of it with the results of 
Ref.~\cite{MEOP2} should allow to test the degree of accuracy of the adopted 
approximations.
To do this the conditions chosen there have been exactly replied: 
a Gaussian initial state with width $\sigma=5$ (in the unit
system in which $\hbar=2m=1$) centered at the origin, an instrumental error 
$\Delta a=1$ and a
measurement time $\tau\simeq 10^{-5}T$. The investigation has been restricted 
to sequences of measurements with constant result $a_n\equiv a_0$ for $n\leq 
N-1$. 
In any symmetrical single-well potential, a symmetrical state localized in 
the middle of the
well will change its width during the dynamical evolution but will remain
centered in the same position at all times. Therefore, starting from such an 
initial state and choosing $a_0=0$, the most probable results of
a measurement is still $\tilde{a}_n\equiv 0$ for every $n$ and any $\Delta T$.
With these values of the parameters, the calculation has been repeated, 
varying the value of $\Delta T$, according to three methods:
\begin{itemize}
\item[${\bf \cal A}$)]
If an impulsive measurement is simulated \cite{MEOP2} by means of a continuous 
one of short duration $\tau\ll\tau_c\stackrel{\rm def}{=}(m/\hbar)
(\Delta a^{-2}+\sigma^{-2})^{-1}$, the computed quantities do not depend upon 
the duration of the measurement.
With the choices made above, during the measurement the state remains a 
Gaussian one, having width
\begin{equation}
\sigma(t)=-\frac
14\left[\alpha+\frac{\beta^2}{4\left(\frac{1}{2\sigma^2}-\alpha\right)}
\right]^{-2},
\end{equation}
with 
\begin{equation}
\alpha=\frac{im\omega_r\cos\omega_r\tau}{2\hbar\sin\omega_r\tau},\qquad
\beta=-\frac{im\omega_r}{\hbar}{\sin\omega_r\tau},\qquad 
\omega_r^2\equiv\omega^2-\frac{i\hbar}{\tau m\Delta a^2}.
\end{equation}
After the end of the measurement, the state evolves causally, still preserving 
the Gaussian form with
\begin{equation}
\sigma(t+\tau)=-\frac
14\left[\alpha'+\frac{\beta'^2}{4\left(\frac{1}{2\sigma^2}-\alpha'\right)}
\right]^{-2};
\end{equation} 
the coefficients refer now to the unmeasured dynamics:
\begin{equation}
\alpha'=\frac{im\omega\cos\omega t}{2\hbar\sin\omega t},\qquad
\beta'=-\frac{im\omega}{\hbar}{\sin\omega t}.\qquad 
\end{equation}
By alternating these two types of evolution and by iterating this procedure 
$N$ times, one can obtain the width $\sigma_N$ of the state after $N$ 
stroboscopic measurements.
In the impulsive regime, $\Delta a_{\rm eff}$ is then easily calculated,
according to its physical meaning \cite{MEOP2}:
\begin{equation}
\label{DAEFFIMP}
\lim_{\tau\to 0}\Delta a_{\rm eff}(\tau)=\sqrt{\Delta a^2+\sigma_N^2}.
\end{equation}
\item[${\bf \cal B}$)]
As shown in Ref.~\cite{MEOP1},  the evaluation of the path integral (\ref{PROPK}) 
can be overcome by writing an effective Schr\"odinger  
equation which takes into account the effect of the measurement through 
an imaginary potential term: 
\begin{equation}
i\hbar\frac{\partial}{\partial t}\psi(x,t)=
\left\{-\frac{\hbar^2}{2m}\frac{\partial^2}{\partial x^2}+V(x)+
\frac{i\hbar}{\tau\Delta a^2}[x-a(t)]^2\right\}\psi(x,t).
\label{schreff}
\end{equation}
Eqn.~(\ref{schreff}) can be resolved with usual numerical techniques
\cite{NUMREC} on a space-time lattice. By turning on and off the measurement 
potential with periodicity $\Delta T$, one can compute 
the values of $\psi(x)$ on the chosen lattice after the desired  
measuring sequence; $\Delta a_{\rm eff}$ is then calculated via
Eqs.~(\ref{PGEN}) and (\ref{DAEFFTH}).
\item[${\bf \cal C}$)]
Method ${\bf \cal A}$ applies only to Gaussian weight functionals.
On the other hand, Eqn.~(\ref{DPHIEFF}) do not need this requirement; 
furthermore, it gives directly the value of $\Delta a_{\rm eff}$
after an arbitrary series of measurements, provided that convenient
approximations have been made. 
The results shown in this paper have been obtained by restricting the extremes 
of integration in Eqn.~(\ref{DPHIEFF}) to $\pm 10\Delta a_{\rm eff}$,
and by truncating the sums to $N_{\rm MAX}=20$. After comparison with the other 
approaches, this method will be applied to forms of the weight function 
different from the Gaussian one (\ref{wepsi}).
\end{itemize}
Fig.~1 shows the typical evolution of the dispersion $\Delta a_{\rm eff}$ 
of the probability $P(a)$, as a consequence of repeated measurements 
with different periodicity (the cases of $\Delta T/T=\frac14$, $\frac 12$ and 
$\frac 34$ are shown).
Suddenly after the first measurement, performed at the initial instant of time, 
$\Delta a_{\rm eff}$ approximates the width of the initial state;
to be more precise, its value is given by Eqn.~(\ref{DAEFFIMP}): in the
considered case it is $\Delta a_{\rm eff}\simeq 5.099$, as evidenced by the 
point $n=1$ of the graph.
Due to the following measurements,  the effective uncertainty reaches an 
asymptotic value $\Delta a_{\rm eff}^{\it as}$ that depends on the quiescent 
time $\Delta T$. The values of $\Delta a_{\rm eff}^{\it as}$ given by the
different methods of calculation show a good agreement: so the more
flexible approach ${\bf \cal C}$ can be held for tested and 
can finally be applied to the comparison between the two measurement theories of
Mensky and of von Neumann, which is the aim of the present paper. This is shown
in the last figures.\\
Fig.~2 depicts
the dependence of the asymptotic probability distribution for the measurement
outcomes, $P_{\it as}(a)$, on the quiescent time. In other words, it is
considered the probability distribution for the result $a$ of the $N^{\rm th}$
measurement ($N=16$ has been chosen, {\it i.e.} in the asymptotic region
as shown in Fig.~1): in the graph are visualized together many
curves relative to different sequences of measurements, each one characterized
by a quiescent time
$\Delta T$, as written on the $y$-axis. It can be seen with great
evidence how the cases with $\Delta T$ integer or semi-integer multiple of the 
oscillation period are characterized by minimal dispersion in the 
$P_{\it as}(a)$.\\
These results are summarized in Fig.~3, which shows the behavior of the 
asymptotic $\Delta a_{\rm eff}$ versus $\Delta T$. It is obtained a periodic
behavior, with minima each 
$\frac{T}{2}$ in which $\Delta a_{\rm eff}^{\it as}$ approximates 
the value $\Delta a$, which identify the optimal QND strategies.
The same pattern would be obtained by choosing $a_0\not=0$, although in this 
case, in general, one has also $\tilde{a}_n\not=a_0$ unless $\Delta T/T=i\in 
{\bf N}$. As explicitly shown above, the latter condition follows from the fact 
that, starting
from a state not centered in the origin, one must wait at least an oscillation
period for the wavepacket to be reformed at the same position.
Thus the definition of $\Delta a_{\rm eff}$ is no longer the same for all
$\Delta T$, except for $\Delta T=T,2T,\ldots$ : for these quiescent times the 
QND strategies are recovered.\\
Finally, the basical aim of Fig.~3 is the comparison between the
results obtained when the 
measurement is simulated respectively by a Gaussian weight (as for Fig.~2)
or by a double-step perfect filter as $\hat{w}^{v.N.}_a$ (\ref{misvn}). 
The most significant difference between the two profiles consists 
in the presence, in the case {\em \`a~la} von Neumann, of oscillations 
and flexes, probably attributable to the discontinuity of $w^{v.N.}_a(x)$. But
the essential qualitative features ({\em i.e.} the ones that
can find experimental application \cite{CAVES}) does not depend upon the 
detailed choice of the weight $w_a$. Thus, for impulsive measurements, von
Neumann wavefunction collapse is simply a particular case of the restricted
path integral theory.

\end{section}

\begin{section}{Conclusions}

The restricted path integral approach to quantum measurement can be 
straightforwardly specialized to impulsive measurements, providing 
analytical formulae for the evolution of the measured system. The model leads
to a filtered state centered around the measurement result through 
instantaneous collapse of the wavefunction, having as width the instrumental 
error.
Von Neumann ideal collapse is recovered by choosing a double-step ideal 
filter as measurement weight functional. The probability 
distribution of the possible measurement outputs has a dispersion whose 
behavior does not depend upon the form of the filter. In particular, for
stroboscopic measurements of position on a harmonic oscillator, the optimal
measuring strategies can be characterized both using an ideal or a Gaussian
filter. On the other hand a continuous, smooth weight appears 
more adequate to represent in simple terms a physical process as the 
measurement. Indeed, the instrument being 
constituted by a collection of microscopic objects immersed in a 
thermodynamical environment \cite{KMN}, it is more likely to give rise to a 
Gaussian probability distribution for the measured physical quantities. A
Gaussian weight is also favorable from the analytical point of view, and
guarantees some natural group properties of the
evolution under continuous monitoring \cite{ME}.\\
Summarizing the obtained results, the restricted path integral theory of 
quantum measurement reveals to be a simple and natural extension of the 
na\"{\i}ve von Neumann instantaneous collapse theory to the more realistic case
of continuous and non-ideal measurements, and appears as a more subtle tool for
handling new fundamental problems (see for instance
Refs.~\cite{MEOP2,OPTA,CALARCO}).

\end{section}
%
%%%%%%%%%%%%%%%%%%%%%%%%%%%ACKNOWLEDGEMENTS
%
\acknowledgements
I am grateful to R.~Onofrio for encouraging me to write this paper, and to him 
and C.~Presilla for a critical reading of the manuscript.
%
%%%%%%%%%%%%%%%%%%%%%%%%%%%%%%%%% REFERENCES LIST
%

%
%%%%%%%%%%%%%%%%%%%%%%%%%%%%%%%%%% FIGURES
%
\begin{figure}[hc]
\caption{Evolution of $\Delta a_{\rm eff}$ versus the number of
measurements $n$, for some values of the quiescent time $\Delta T$.
The continuous line refers to the analytical calculation done in 
Ref.~\protect\cite{MEOP2} (crosses are numerical results) by approximating each 
impulsive measurement with a continuous one of short duration $\tau\simeq 
10^{-5}T$. The dotted line is deduced from Eqn.~(\protect\ref{DPHIEFF}), 
{\em i.e.} in the limit for $\tau\to 0$. 
The agreement, besides slight deviations in the initial transient, is within 
1\% in the asymptotic region.}
\end{figure}

\begin{figure}[c]
\caption{Asymptotic probability distribution $P_{\it as}(a)$ for the
measurement outcomes $a$ as function of the quiescent time 
$\Delta T/T$. The curve reaches its maxima (and correspondingly the
minimum dispersion in $a$) at $\Delta T$ multiples of half an
oscillation period. On top of the graph is superimposed a contour plot which
shows the extremal regions.}
\end{figure}

\begin{figure}[hc]
\caption{Normalized asymptotic effective uncertainty $\Delta a_{\rm eff}^{\it
as}/\Delta a$ versus the quiescent time $\Delta T/T$ for stroboscopic
measurements performed with Gaussian (\protect\ref{PESOW}) or perfect
(\protect\ref{wepsi}) filtering. The curves does not differ qualitatively and
coincide at the minima, as theoretically expected \protect\cite{MEOP2}.}
\end{figure}

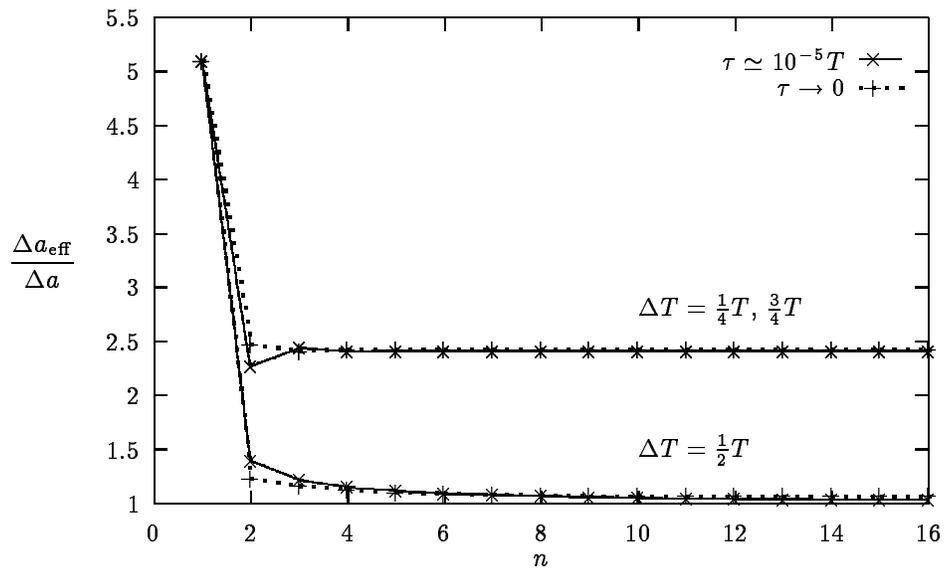
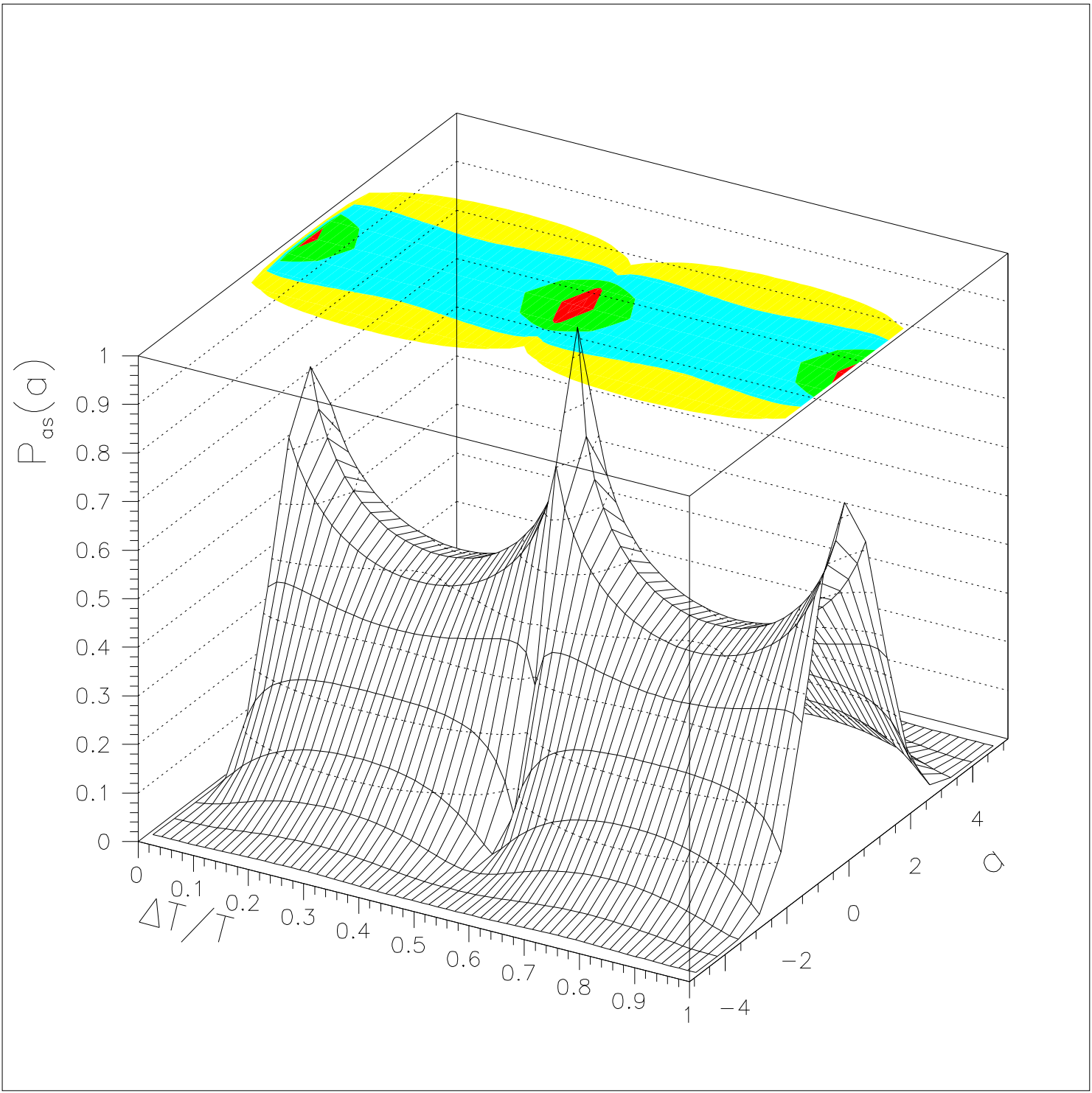
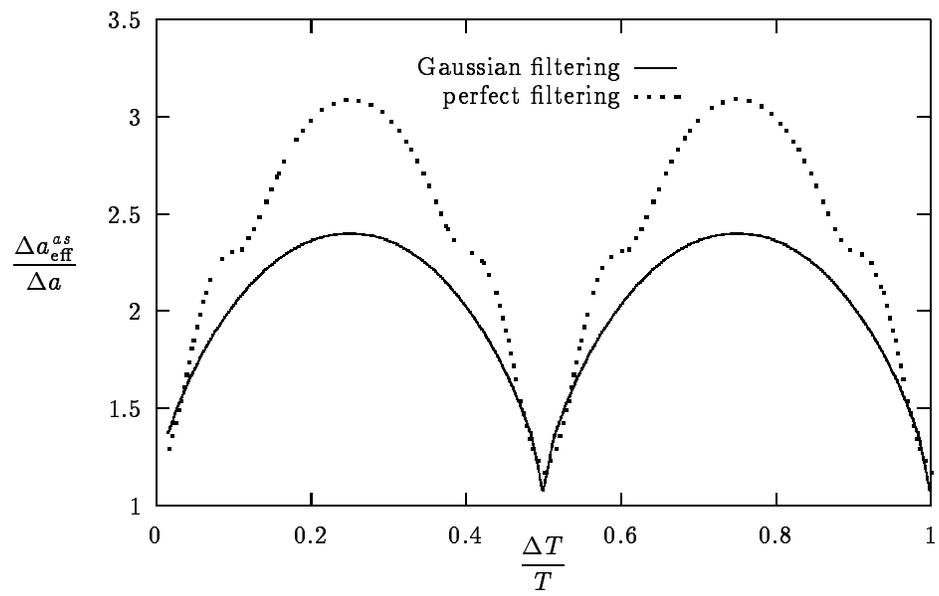
\end{document}